\documentclass[twocolumn,showpacs,preprintnumbers,unsortedaddress,superscriptaddress,floatfix]{revtex4}
\usepackage{amsmath}
\usepackage{graphicx}
\usepackage{dcolumn}
\usepackage{bm}
\usepackage{amsfonts}
\usepackage{amssymb}
\usepackage{mathrsfs}
\usepackage{subfigure}

\begin{document}

\title{Experimental Measurement of Lower and Upper Bounds of Concurrence for
Mixed Quantum States}
\author{Xiao-Ling Niu}
\affiliation{Key Laboratory of Quantum Information, University of Science and Technology
of China, CAS, Hefei 230026, People's Republic of China}
\author{Yan-Xiao Gong}
\affiliation{Key Laboratory of Quantum Information, University of Science and Technology
of China, CAS, Hefei 230026, People's Republic of China}
\author{Jian Li}
\affiliation{Physics Department of SouthEast University}
\author{Liang Peng}
\affiliation{Key Laboratory of Quantum Information, University of Science and Technology
of China, CAS, Hefei 230026, People's Republic of China}
\author{Cheng-Jie Zhang}
\affiliation{Key Laboratory of Quantum Information, University of Science and Technology
of China, CAS, Hefei 230026, People's Republic of China}
\author{Yong-Sheng Zhang}
\affiliation{Key Laboratory of Quantum Information, University of Science and Technology
of China, CAS, Hefei 230026, People's Republic of China}
\author{Yun-Feng Huang}
\email{hyf@ustc.edu.cn}
\affiliation{Key Laboratory of Quantum Information, University of Science and Technology
of China, CAS, Hefei 230026, People's Republic of China}
\author{Guang-Can Guo}
\affiliation{Key Laboratory of Quantum Information, University of Science and Technology
of China, CAS, Hefei 230026, People's Republic of China}

\begin{abstract}
We experimentally measure the lower and upper bounds of concurrence for a
set of two-qubit mixed quantum states using photonic systems. The measured
concurrence bounds are in agreement with the results evaluated from the
density matrices reconstructed through quantum state tomography. In our
experiment, we propose and demonstrate a simple method to provide two
faithful copies of a two-photon mixed state required for parity
measurements: Two photon pairs generated by two neighboring pump laser
pulses through optical parametric down conversion processes represent two
identical copies. This method can be conveniently generalized for
entanglement estimation of multi-photon mixed states.
\end{abstract}

\maketitle

Quantum entanglement plays a key role in not only fundamental quantum
physics but also quantum information processing. Consequently the
characterization and quantification of entanglement have attracted much
attention \cite{1,2,3,4,5,6} and various entanglement measures have been
proposed, such as concurrence \cite{2,7,8}, negativity \cite{4} and tangle
\cite{5}. With these theoretical progresses, experimental quantification of
entanglement becomes a natural requirement. However, it is a rather
difficult task, since many entanglement measures proposed are complicated
nonlinear functions of the density matrix of the quantum state. One simple
method to estimate entanglement degree is quantum state tomography \cite%
{9,10}, which has been applied successfully in a number of experiments \cite%
{11,12,13}. In quantum state tomography, one measures a complete set of
observables and reconstructs the density matrix of the measured quantum
state, and then the left thing is to mathematically evaluate some
entanglement measure using the density matrix. Since one has to measure a
complete set of observables for tomography, this leads to rapidly growing
experimental overhead as system size increases, either in subsystem
dimensions or in subsystem numbers. That makes quantum state tomography an
unscalable method. Another important method for entanglement detection is
the entanglement witness \cite{14,15,16,17}, which provides a much more
direct experimental insights to the entanglement property of a quantum
state. However, it requires some a priori knowledge on the state to be
detected. So entanglement witnesses can not be freely applied for arbitrary
unknown quantum states.

To overcome the above drawbacks, Mintert \textit{et al.} \cite{18} recently
proposed a method to directly measure the concurrence of an arbitrary pure
state $\left\vert \Psi \right\rangle $ through a single projection
measurement on its twofold copy $\left\vert \Psi \right\rangle \otimes
\left\vert \Psi \right\rangle $. Based on this method, experimental
measurements of concurrence for two-qubit \cite{19,20} and $4\times 4$%
-dimensional \cite{21} pure states have been reported. However, for more
general applications one would like to have the ability to experimentally
measure entanglement of not only pure states, but also mixed states. For
this purpose, Mintert \textit{et al.} \cite{22} and Aolita \textit{et al.}
\cite{23} presented observable lower bounds of concurrence for arbitrary
bipartite and multipartite mixed states, respectively. After that, some of
us \cite{24} presented observable upper bound of concurrence for arbitrary
finite-dimensional mixed states.

In this paper, we report the first experimental measurements of lower and
upper bounds of concurrence for a set of two-qubit mixed states using the
twofold copy parity measurements in \cite{22} and \cite{24}. Our results
give an exact region which must contain the concurrence of the measured
mixed states. We also reconstruct the density matrices of the mixed states
through quantum state tomography and evaluate lower and upper bounds of
concurrence with the density matrices. We find that the experiment results
obtained by these two methods are in agreement with each other. So far most
experiments investigating entanglement properties have invoked photonic
systems \cite{11,12,13,19,21,25} due to their mature manipulation
technologies and wide applications in quantum information science. However
an important experimental difficulty to realize the measurements in \cite{22}
and \cite{24} using photon systems is the preparation of two identical
copies of an unknown mixed state $\rho $, that is, one who wants to
implement these measurements has to be certain of the source providing $\rho
\otimes \rho $, but can be perfectly ignorant about the initialization of
the source, as pointed out by Mintert et al. in \cite{22}. In our
experiment, we present and demonstrate a method by which one can easily
prepare a reliable photon source providing two faithful copies of an unknown
mixed state.

\begin{figure}[tb]
\centering
\includegraphics[width=7cm]{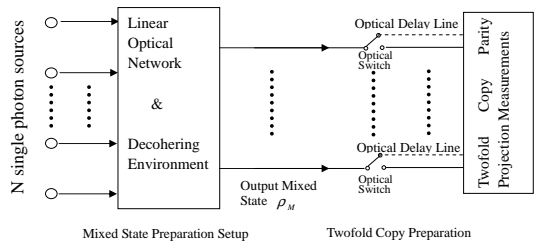}
\caption{Scheme for preparing two faithful copies of an unknown multi-photon
mixed state.}
\label{Fig1}
\end{figure}

Our method utilizes the copies carried by photons emitted out from the same
photon sources and passing through the same preparation setup, but at
different times. The method is depicted in Fig.~\ref{Fig1}. Suppose in the
most general situation of linear optics quantum information processing in
the future, we have $N$ single photon sources which can emit one single
photon pulse at one time. After passing through some linear optical quantum
computing networks and at the same time suffering from some decoherence
processes, the output $M$ photons will be in a multi-partite mixed state $%
\rho _{M}$. Now if we want to detect the entanglement of this multi-photon
mixed state, we can send each photon into an optical delay line (This can be
achieved by sending photons into new optical paths and controlling the
length of these new paths as delay line). Then let's wait for the
multi-photon state generated by next pulses of these single photon sources.
Since all photons pass through exactly the same state preparation setup,
this multi-photon state must be in the same mixed state $\rho _{M}$. Thus we
have got two faithful copies of the same multi-photon mixed state. Moreover,
we even don't need to control the state preparation setup. That means this
method can provide us two identical copies of any unknown multi-partite
mixed state. At last we can send the photons in the new paths and original
paths into the parity measurements setup for concurrence bounds measurements
as proposed in \cite{23}.

\bigskip Now let us briefly introduce the measurements in \cite{22} and \cite%
{24}. The lower bound of concurrence for an bipartite mixed state presented
in \cite{22} is described by the following inequality

\bigskip
\begin{equation}
\left[ C(\rho )\right] ^{2}\geq Tr(\rho \otimes \rho V_{i})\ \ \ \ \ \
(i=1,2).  \label{equation1}
\end{equation}%
Here $V_{1}=4(P_{-}-P_{+})\otimes P_{-}$ and $V_{2}=4P_{-}\otimes
(P_{-}-P_{+})$. $P_{-}$ is the projector on the antisymmetric subspace of
the two copies of either subsystem and $P_{+}$ is the symmetric counterpart
of $P_{-}$. The upper bound derived in \cite{24} also corresponds to the
inequality

\begin{equation}
\left[ C(\rho )\right] ^{2}\leq Tr(\rho \otimes \rho K_{i})\ \ \ \ \ \
(i=1,2),  \label{equation2}
\end{equation}%
where $K_{1}=4(P_{-}+P_{+})\otimes P_{-}$, $K_{2}=4P_{-}\otimes
(P_{-}+P_{+}) $. However, for the case of two-qubit mixed states, there is a
tighter upper bound \cite{24}

\begin{equation}
\left[ C(\rho )\right] ^{2}\leq Tr(\rho \otimes \rho \cdot 4P_{-}\otimes
P_{-})\   \label{equation3}
\end{equation}%
Thus in experiment we only need to make a few parity projection measurements
$P_{-}\otimes P_{-}$, $P_{-}\otimes P_{+}$ and $P_{+}\otimes P_{-}$ on the
twofold copy to evaluate the lower and upper bounds.

\begin{figure}[tb]
\centering
\includegraphics[width=7cm]{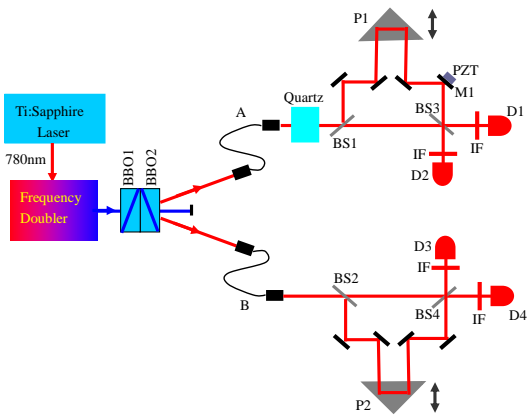}
\caption{(Color online) Experiment setup for measuring lower and upper
bounds of concurrence for two-photon mixed states. IF is interference filter
centered at $780$ $nm$ wavelength with $3$ $nm$ bandwidth.}
\label{Fig2}
\end{figure}

Our experiment setup is outlined in Fig.~\ref{Fig2}. Instead of using single
photon sources, here we use photon pairs produced by optical parametric
down-conversion processes to prepare two-qubit mixed states. A pulse train
from a mode-locked Ti:sapphire laser (with a duration of $140$ $fs$, a
repetition rate of $76$ $MHz$ and a central wavelength of $780$ $nm$) first
passes through a frequency doubler. Then the ultraviolet pulses (in $H$
polarization, with $200$ $mW$ average power) from the doubler pump two $1$ $%
mm$ thick $\beta $-barium borate crystals (BBO1 and BBO2) located side by
side to generate polarization-entangled photon pairs. The performance and
detailed description of this photon pair source can be found in \cite{26}.
With this source, the output photon pairs from single-mode fibers(A and B)
are in the maximally entangled state $\frac{1}{\sqrt{2}}(\left\vert
HV\right\rangle -\left\vert VH\right\rangle )$. Then one photon of the pair
(from single-mode fiber A) passes through a phase-damping channel which is
composed of a birefringent crystal (we use quartz crystal here). After that
the twin-photon is prepared in a certain two-qubit mixed state. Thus by
changing the thickness of the quartz crystal we can prepare a set of
two-qubit mixed state with different concurrence values. Next step is to
prepare a twofold copy of this mixed state. For experiment convenience, we
use two 50/50 beamsplitters (BS1 and BS2) in our experiment instead of the
optical switches in Fig. 1. The only drawback of this change is the decrease
of total detection efficiency. At last another two 50/50 beamsplitters (BS3
and BS4) are used to make the $P_{-}$ projection measurement on the twofold
copy of each subsystem. So the coincidence counts of single photon detectors
D1, D2, D3, and D4 correspond to the result of $P_{-}\otimes P_{-}$
measurement. To assure BS3 (BS4) making $\ P_{-}$ measurement, the optical
path lengths between BS1 (BS2) and BS3 (BS4) are carefully arranged so that
the reflected photon on BS1 (BS2) can arrive at BS3 (BS4) at the same time
with the transmitted photon generated by the next pump laser pulse. In the
experiment the reflection arm is about $3.947$ meters longer than the
transmission arm. This length corresponds to the spatial distance between
two neighbouring pump laser pulses, which is $c\ast \frac{1}{76\text{ }MHz}$%
, with $c$ denoting the speed of light. If the two photons have the same
polarization, by moving the right angle prism P1 (P2) mounted on a
translational stage and observing the two-photon coincidence counts of D1
and D2 (D3 and D4), we can observe a two-photon interference dip, with
theoretical visibility 1/3. The interference of these two photons is
intrinsically the same as the interference between two photons from two
spatially separated photon pair sources \cite{27}. In our experiment, the
visibilities of two interferences at BS3 and BS4 are both $0.30$. Now if we
make the two translational stages stay exactly at the dip places, the
four-photon coincidences should correspond to the result of $P_{-}\otimes
P_{-}$ measurement.

According to the original scheme in \cite{22}, we should also perform $%
P_{-}\otimes P_{+}$ and $P_{+}\otimes P_{-}$ measurements. Here we use a
little trick for experimental convenience. Notice that $P_{-}+P_{+}=I$, so
we can perform $P_{-}\otimes I$, $I\otimes P_{-}$ measurements to evaluate $%
P_{-}\otimes P_{+}$ and $P_{+}\otimes P_{-}$. We also need to perform $%
I\otimes I$ measurement for normalization. The $I$ measurement on the
twofold copy of either subsystem in our experiment setup can be easily
realized by moving the translational stage out of the two-photon
interference region. In this case, the measured two-photon coincidences of
D1 and D2 (D3 and D4) should correspond to $\frac{1}{2}I$ measurement. In
this way we can get the results of $\frac{1}{2}I\otimes P_{-}$, $%
P_{-}\otimes \frac{1}{2}I$ and $\frac{1}{2}I\otimes \frac{1}{2}I$
measurements. Combining the measured result of $P_{-}\otimes P_{-}$, we can
evaluate the lower and upper bounds of concurrence.

\bigskip In the above analysis, we only consider the case that the first and
second pump laser pulses each generates one photon pair. But due to the
specialty of spontaneous parametric down conversion source, there is equal
probability that two photon pairs are produced by only one pump pulse (the
first or the second). These two additional cases also have contributions to
four-photon coincidences. So we have to subtract these backgrounds from the
above measurement results. To do this, we block the reflection arm of one
subsystem and record the four-photon coincidences as $b1$. Similarly, the
four-photon coincidences when blocking the transmitted arm of one subsystem
are recorded as $b2$. Then the background coincidences are $b1+b2$. Notice
that these backgrounds keep constant against the location of translational
stages because the contributions from the additional two cases have no
relationship with two-photon interferences on BS3 and BS4. So we can
subtract $b1+b2$ from each measured coincidence counts and finally obtain
the net results of $P_{-}\otimes P_{-},\frac{1}{2}I\otimes P_{-}$, $%
P_{-}\otimes \frac{1}{2}I$ and $\frac{1}{2}I\otimes \frac{1}{2}I$.

Another problem should be mentioned is the interference effect between the
above three cases. Due to the mode-locked property of the pump laser, the
phase of one pump laser pulse is locked with the next pulse. Thus the
four-photon coincidences contributions from these three cases should be
coherent in phase. Such phase coherence would cause interference effect on
four-photon coincidences and could spoil experiment results. To remove this
effect, we mount one reflection mirror (M1) on a piezoelectric transducer
(PZT) and drive the PZT with a random voltage. This will induce a random
phase between each of the three cases and thus destroy the phase coherence
of them \cite{28}. In our experiment, due to a very long time required for
observing four-photon coincidences, we observe the two-photon coincidences
of D1 and D3 instead to make sure that phase coherence has been destroyed.
The two-photon coincidences may come from two cases: the first pump pulse
generates one photon pair and the photon pair passes through the reflection
arm; or the second pulse generates one pair while the photon pair passes
through the transmitted arm. These two cases have similar phase coherence
with the above three cases. By observing coincidences of D1 and D3, we find
that the coincidences varies obviously in the time scale of a few seconds
when no voltage is applied on the PZT. This is because the relative phase
between the reflection and transmission arm is not stable, as in normal
Mach-Zehnder interferometers. But when we drive the PZT with an random
signal, the coincidences become stable and no interference effect can be
observed. This phenomenon demonstrates that phase coherence between such
cases can be effectively destroyed using this method.

\begin{table}[tbp]
\caption{Experiment results for eight mixed states via two methods. $%
C_{l}^{twofold}$ and $C_{u}^{twofold}$ are lower and upper bounds of
concurrence obtained by twofold copy parity projection measurements,
respectively $C^{tom}$, $C_{l}^{tom}$, and $C_{u}^{tom}$ are concurrence and
its lower and upper bounds evaluated by tomography, respectively. $T_{quartz%
\text{ }}$is thickness of the quartz crystal for corresponding mixed state.}
\label{TableI}%
\begin{ruledtabular}
\begin{tabular}{llllll}
$T_{quartz}$ $(mm)$ & $C_{l}^{tom}$ & $C_{l}^{twofold}$ & $C^{tom}$ & $%
C_{u}^{twofold}$ & $C_{u}^{tom}$ \\
0 & 0.931 & 0.860$\pm 0.063$ & 0.932 & 0.949$\pm 0.027$ & 0.965 \\
2.985 & 0.908 & 0.801$\pm 0.086$ & 0.910 & 0.869$\pm 0.035$ & 0.955 \\
6.584 & 0.812 & 0.611$\pm 0.071$ & 0.815 & 0.812$\pm 0.024$ & 0.910 \\
9.568 & 0.669 & 0.705$\pm 0.084$ & 0.672 & 0.877$\pm 0.031$ & 0.851 \\
13.167 & 0.539 & 0.388$\pm 0.142$ & 0.539 & 0.833$\pm 0.029$ & 0.803 \\
17.468 & 0.349 & 0.297$\pm 0.158$ & 0.376 & 0.686$\pm 0.029$ & 0.747 \\
20.453 & 0.237 & 0.250$\pm 0.213$ & 0.239 & 0.835$\pm 0.029$ & 0.727 \\
24.052 & 0.00 & 0.182$\pm 0.234$ & 0.092 & 0.782$\pm 0.024$ & 0.703%
\end{tabular}
\end{ruledtabular}
\end{table}

The experiment results are listed in Table~\ref{TableI}. We measured the
concurrence bounds of eight mixed states. Eight quartz crystals with
different thicknesses ranging from $0$ to $24$ $mm$ are employed as
decohering environment to prepare these mixed states. From Table I we can
see that most lower and upper bounds calculated from parity projection
measurements are compatible with the results evaluated through tomography
method, considering the error bars caused by photon counting statistics.
Furthermore, for most states the concurrence evaluated by tomography are in
the region between the lower and upper bounds obtained from parity
projection measurments. Here we didn't calculate the density matrices of
these mixed states from the parameters of our experiment setup, since state
preparation is not the purpose of this experiment. Instead we compare the
results of parity projection measurements with the results via tomography,
because quantum state tomography has been applied rather successfully for
two-qubit cases \cite{9}.

Comparing these two methods in such two-qubit case, it seems that the method
of twofold copy parity measurements is more complicated and needs more data
collection time, since two two-photon interferometers are used and
four-photon coincidence counts are recorded as experiment data. Furthermore,
in our experiment we need to subtract about $\frac{2}{3}$ coincidence counts
as backgrounds, which makes data collection time even longer. However, if we
consider more general case of many photonic qubits and spontaneous
parametric down-conversion sources being replaced by single photon sources
in the future, as shown in Fig.~\ref{Fig1}, the number of two-photon
interferometers only increases linearly with qubit numbers and data
collection time would not increase exponentially either. On the other hand,
the exponentially increasing experiment resources for tomography is
inevitable. That makes our method more suitable for multi-photon case than
quantum state tomography. In this context, it is meaningful for us to give a
proof-in-principle experimental demonstration.

In summary, we for the first time experimentally measured the lower and
upper bounds of concurrence for a set of two-qubit mixed states using
twofold copy parity projection measurement method. The measured results are
compatible with the results evaluated from conventional quantum state
tomography. The technique we used to provide two faithful copies of an
unknown mixed states is perfectly ignorant of the specific mixed state $\rho
$ and can be easily generalized for many photonic qubits case in the future.
This might be helpful for research on quantum entanglement property of
multi-qubit systems.

We thank Zhe-Yu Ou for helpful discussion. This work was funded by National
Fundamental Research Program (grant 2006CB921907); National Natural Science
Foundation of China (Grants No. 10674127, No. 60621064 and No. 10774139),
Innovation Funds from Chinese Academy of Sciences, Program for New Century
Excellent Talents in University; International Cooperation Program from CAS
and Ministry of Science and Technology of China; and A Foundation for the
Author of National Excellent Doctoral Dissertation of PR China (grant
200729).

\bigskip

\bigskip

\bigskip

\bigskip

\bigskip

\end{document}